\documentclass[aps,prc,twocolumn,times,graphicx,tighten,showpacs]{revtex4}
\usepackage[dvips]{graphicx}

\newcommand{\bea}{\begin{eqnarray}}
\newcommand{\eea}{\end{eqnarray}}
\newcommand{\be}{\begin{equation}}
\newcommand{\ee}{\end{equation}}

\begin{document}

\title{
Meson-exchange currents and final-state interactions
in quasielastic electron scattering at high momentum transfers
}

\author{
J.E. Amaro$^a$, 
M.B. Barbaro$^b$,
J.A. Caballero$^c$, 
T.W. Donnelly$^d$,
C. Maieron$^a$,
J.M. Udias$^e$
} 

\affiliation{$^a$Departamento de Fisica Atomica, Molecular y Nuclear, 
Universidad de Granada,
 Granada 18071, Spain}

\affiliation{$^b$Dipartimento di Fisica Teorica, Universit\`a di Torino and
  INFN, Sezione di Torino, Via P. Giuria 1, 10125 Torino, Italy}

\affiliation{$^c$Departamento de F\'{\i}sica At\'omica, Molecular y Nuclear, 
Universidad de Sevilla, Apdo.1065, 41080 Sevilla, Spain}

\affiliation{$^d$Center for Theoretical Physics, Laboratory for Nuclear
  Science and Department of Physics, Massachusetts Institute of Technology,
  Cambridge, MA 02139, USA}

\affiliation{$^e$Departamento de F\'{\i}sica At\'omica, Molecular y Nuclear, 
Universidad Complutense de Madrid, 28040, Madrid, Spain}

\date{\today}


\begin{abstract}
  The effects of meson-exchange currents (MEC) are computed for the
  one-particle one-hole transverse response function for finite nuclei
  at high momentum transfers $q$ in the region of the quasielastic
  peak.  A semi-relativistic shell model is used for the
  one-particle-emission $(e,e')$ reaction. Relativistic effects are
  included using relativistic kinematics, performing a
  semi-relativistic expansion of the current operators and using the
  Dirac-equation-based (DEB) form of the relativistic mean field
  potential for the final states.  It is found that final-state
  interactions (FSI) produce an important enhancement of the MEC in
  the high-energy tail of the response function for $q\geq 1$ GeV/c.
  The combined effect of MEC and FSI goes away when other models of
  the FSI, not based on the DEB potential, are employed.
\end{abstract}

\pacs{25.30.Fj; 21.60.Cs; 24.10.Jv}

\maketitle

In recent years much of the emphasis in studies of inclusive $(e,e')$
scattering has been placed on investigations of the scaling properties
of the cross section and on the possibility of predicting neutrino
cross sections assuming the universality of the scaling function for
electromagnetic and weak interactions. An exhaustive analysis of
$(e,e')$ world data has demonstrated the scaling at energy transfers
$\omega$ below the QE peak\cite{DS199,DS299}, namely the independence
of the reduced cross sections on the momentum transfer (first-kind
scaling) and on the nuclear target (second-kind scaling) when plotted
versus the appropriate scaling variable.  It is well known that at
energies above the QE peak scaling is violated in the transverse (T)
channel by effects beyond the impulse approximation: inelastic
scattering~\cite{Alvarez-Ruso:2003gj,BCDM04}, correlations and MEC in
both the one-particle one-hole (1p-1h) and two-particle two-hole
(2p-2h) sectors~\cite{Amaro:2001xz,Amaro:2002mj,Amaro:2003yd,DePace}.

In contrast, the available data for the longitudinal (L) response are
compatible with scaling throughout the QE region and have permitted
\cite{MDS02} the extraction of a phenomenological scaling function
$f_L$. In recent work \cite{Caballero:2005sj,Ama06,Barbaro:2008zv} it
has been shown that only a few models (the relativistic mean field
(RMF), the semi-relativistic (SR) approach with DEB and a ``BCS-like''
model) are capable of reproducing the detailed shape of $f_L$, while
other models fail to reproduce the long tail appearing at high
$\omega$.  The above models effectively account for the major
ingredients needed to describe the $(e,e')$ responses for
intermediate-to-high momentum transfers, namely relativitic effects
and an appropriate description of the effective FSI.

Approximate treatments of these two ingredients are also possible
using SR models, which have the advantage of permitting the use of
standard non-relativistic techniques when correctly extrapolated to
high values of $q$.  In this paper we use the approach of
\cite{Ama05,Ama06}, where a specific SR expansion of the electroweak
single-nucleon current was used in a continuum shell-model description
of electron and neutrino inclusive QE scattering from closed-shell
nuclei.  In the model the (non-relativistic) hole states are taken to
be states in a Woods-Saxon potential, while the final particles in the
continuum are described with the DEB form of the RMF plus the
so-called Darwin term. Similar studies have shown that the T response
computed in impulse approximation has the same scaling properties as
the L response. The deviations from scaling observed in data for the T
response are usually ascribed to mechanisms beyond the one-particle
emission channel, specifically, two-particle emission, delta
excitation and other inelastic processes.

\begin{figure}[tph]
\begin{center}
\includegraphics[scale=0.65,  bb= 130 470 490 690]{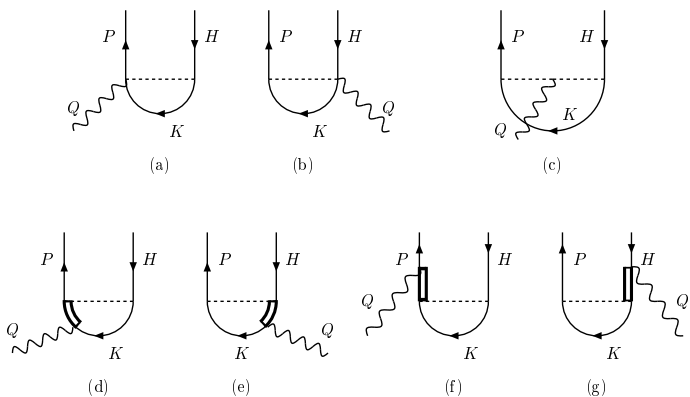}
\caption{
One-particle ($P)$ one-hole ($H$) MEC diagrams considered in the present study.
Diagrams (a,b) correspond to the seagull, (c) to the pionic, and (d-g)
to the $\Delta$ current, respectively.  The intermediate particle $K$ 
corresponds to a sum over occupied holes in the shell-model core.}
\label{Fig1}
\end{center}
\end{figure}

In this paper we focus on a study of the MEC contributions in the
1p-1h transverse QE response at high momentum transfers.  The MEC here
are 2-body contributions which at the 1p-1h level occur coherently
with the familiar 1-body contributions; the former are depicted in the
diagrams of Fig.~1. Most of  MEC studies performed for low-to-intermediate
momentum transfers 
\cite{Amaro:2002mj,Amaro:2003yd,Alberico:1989aja,Koh81,Dek92,Fab97,Slu95}
have shown a small reduction of the total
response at the peak. These are produced mainly via the $\Delta$
current (diagrams (d-g) --- excitation of a virtual $\Delta$ which
subsequently decays, exchanging a pion with a nucleon in the Fermi
sea), while the seagull (S) and pion-in-flight (P) currents (diagrams
(a-b) and (c) respectively) give a net positive but smaller
contribution \cite{Ama94}.  Specifically, the destructive interference
between the familiar 1-body and 2-body contributions yields a 12\%
reduction of the total at $q=500$ MeV/c, rising to about 20\% at $q=1$
GeV/c. The shape of the T response does not change too much for such
kinematics, resulting in only small scaling violations in the total T
channel.  Similar effects have also been found at higher momentum
transfers using a relativistic Fermi gas (RFG) model
\cite{Amaro:2001xz,Amaro:2003yd}.  The same trend is confirmed by the
results of the present study for $q=1$ GeV/c, as shown in the upper
panel of Fig.~2: here the T response obtained using the DEB model with
and without MEC is displayed.

\begin{figure}[tph]
\begin{center}
\includegraphics[scale=0.7,  bb= 150 270 430 770]{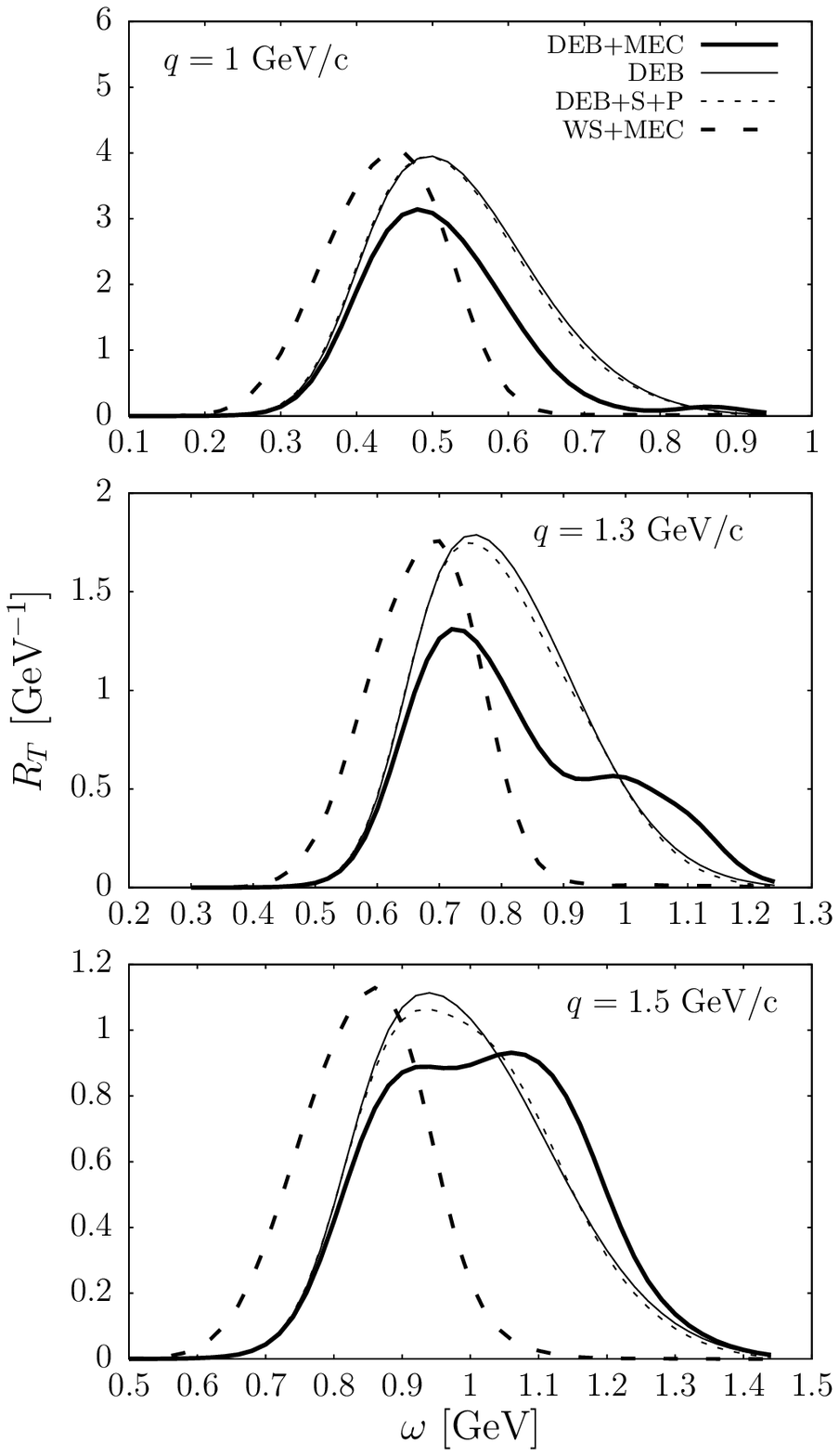}
\caption{
Transverse response for $^{12}$C versus $\omega$ for three values of
$q$. The results with the DEB potential including the full MEC (thick solid
lines), the seagull (S) and pion-in-flight (P) currents (thin-dashed) and only  
the 1-body current (thin solid) are displayed. Also shown are the
results corresponding to the Woods-Saxon potential with the full MEC contribution
(thick-dashed).}
\label{Fig2}
\end{center}
\end{figure}

However, the behavior changes for higher values of $q$, as shown in
the lower panels of Fig.~2. For the DEB+MEC approach the reduction of
the response in the peak region due to MEC is now accompanied by an
increase of the tail in the high-$\omega$ region, where an enhancement
of $R_T$ appears as a bump, producing a drastic change from the usual
QE peak shape.  Therefore one expects a large violation of 1st-kind
scaling in this region even when considering only the 1p-1h T
response.  The amount of violation increases with $q$, and for $q=1.5$
GeV/c a plateau-like shape is obtained.  On the other hand, such
peculiar behavior for high $q$ is not observed when using a
Woods-Saxon (WS) potential for the final states.  The reason is that
the MEC bump appears in the high-energy region where the WS results
are very small, and therefore not observable in the figure, whereas
the DEB potential gives rise to a long tail which emphasizes the
effects occurring at large $\omega$.

\begin{figure}[tph]
\begin{center}
\includegraphics[scale=0.7,  bb= 150 280 430 790]{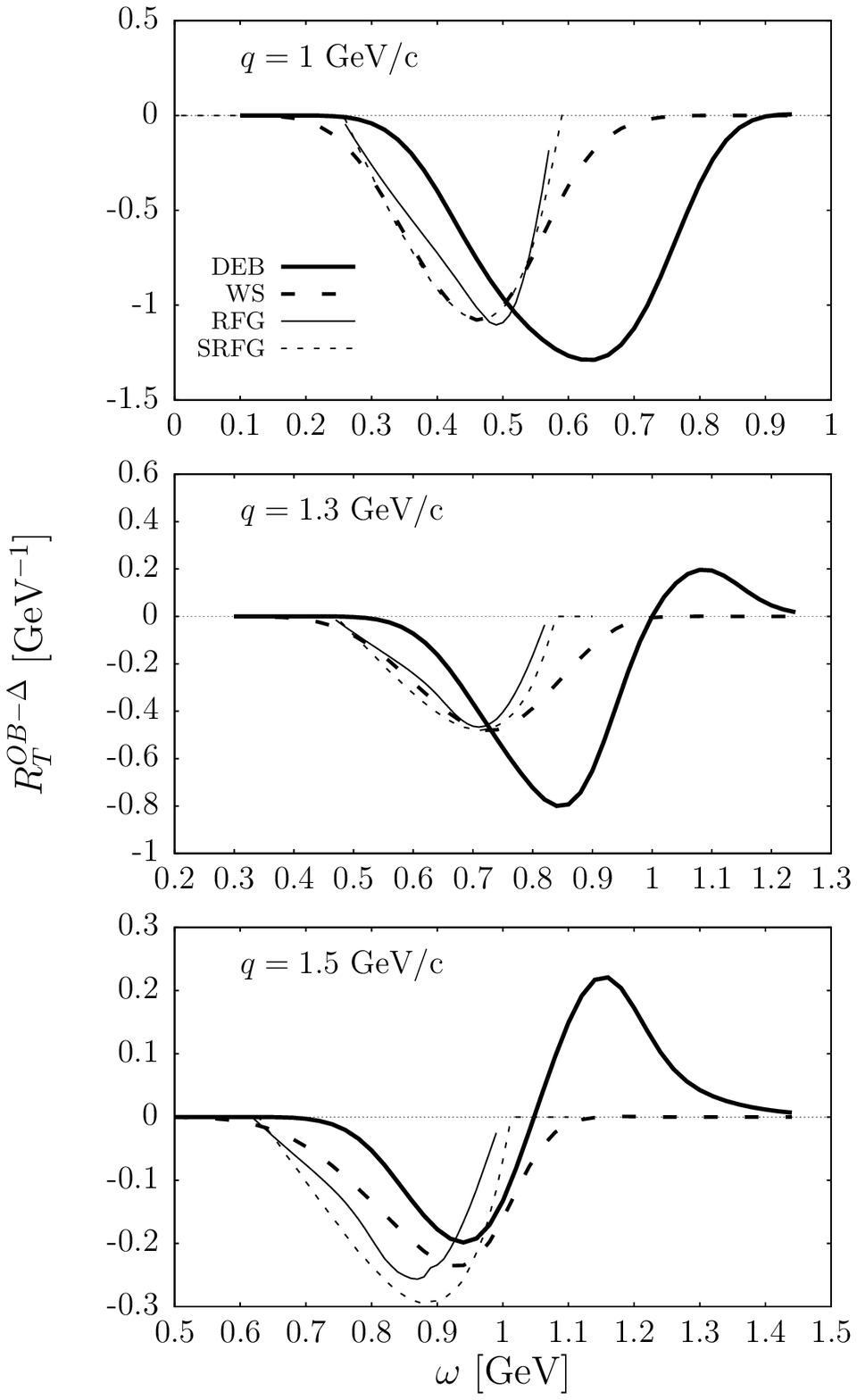}
\caption{ The transverse response corresponding to the interference
  between 1-body and $\Delta$ currents is shown for the DEB
  (thick-solid) and Woods-Saxon (thick-dashed) potentials, the
  Relativistic Fermi Gas (thin-solid) and the SR Fermi
  gas (thin-dashed).}
\label{Fig3}
\end{center}
\end{figure}

 From Fig.~1 it also appears that the seagull and pion-in-flight
 diagrams very weakly affect $R_T$ while the largest MEC contribution
 comes from the interference term between 1-body and 2-body $\Delta$
 currents. The SR $\Delta$ current used in this work is taken from
 \cite{Ama2003} and includes a static propagator for the intermediate
 $\Delta$.  One might think that the static approximation should not
 be adequate for high energies and momentum transfers. However, it was
 shown in \cite{Ama2003} to be the best approximation among several
 prescriptions for the dynamical propagator for $q<1$ GeV/c, when
 comparing with the RFG model of \cite{Amaro:2003yd}, where the
 relativistic $\Delta$ propagator is treated exactly.  For the present
 study we have extended the SR model for the $\Delta$ current to
 $q=1.5$ GeV/c.  We have first checked that at these high values of
 $q$ the static approximation is still valid, as demonstrated in
 Fig.~3 where we show the interference between the $\Delta$ and OB
 current in the T response.  Indeed the SR Fermi gas results are seen
 to be very close to the RFG ones even for $q$ as high as 1.5 GeV.  In
 the same figure we also show the SR shell-model results using the WS
 potential.  The RFG and WS are very similar, except for the
 kinematical region where the RFG is zero. Finally, in Fig.~3 we show
 the results obtained using the DEB potential, which produces a
 significant hardening of the response and an oscillatory behavior at
 high $q$.  A change of sign appears above $\omega \simeq 1$ GeV and
 is responsible for the MEC bump observed in Fig.~2.  By closer
 inspection of Fig.~3, a change of sign can be also observed in the WS
 results for the same energy, although the response is so small in
 that region that the effect is negligible (see above).

\begin{figure}[tph]
\begin{center}
\includegraphics[scale=0.7,  bb= 150 440 420 770]{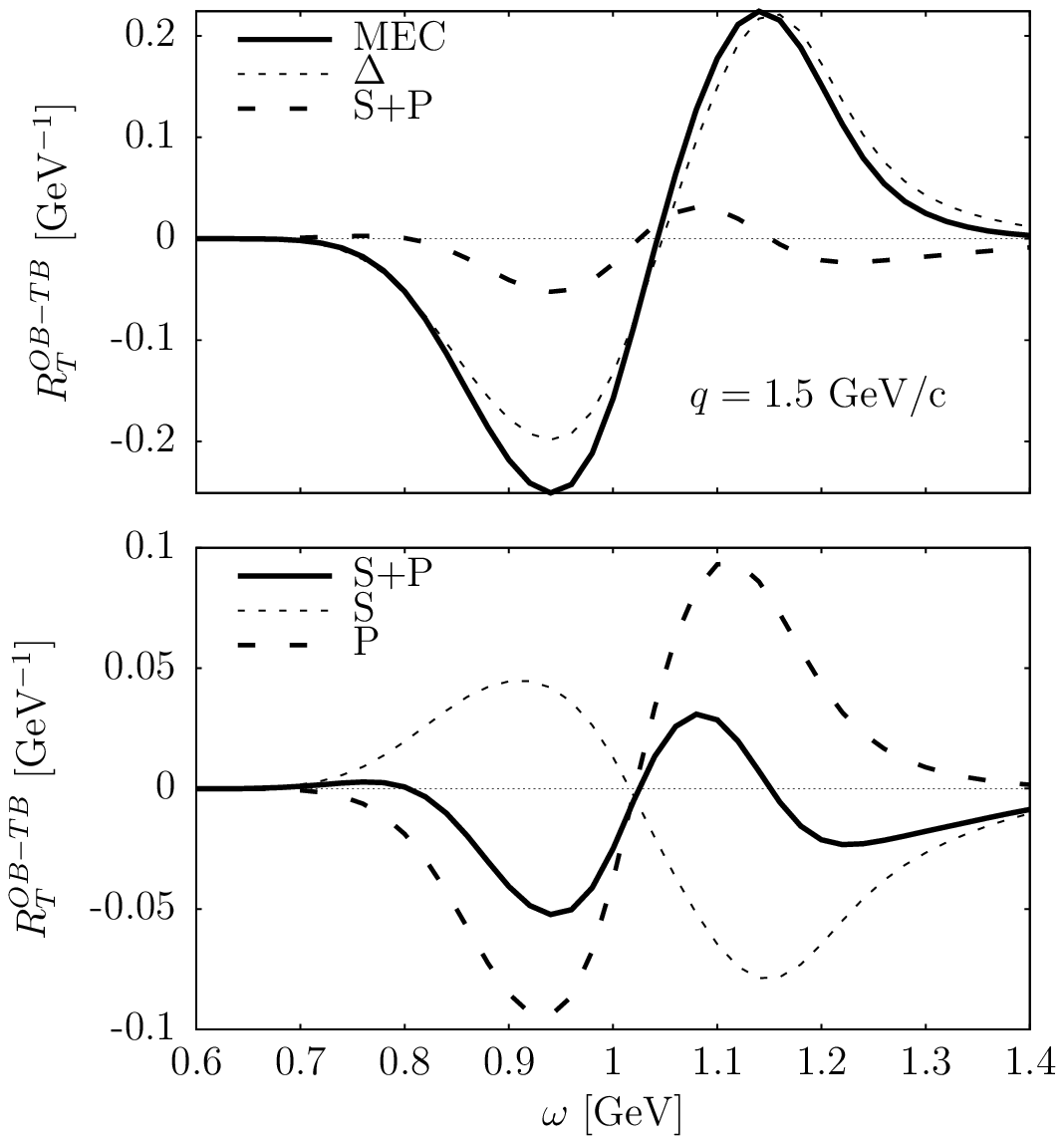}
\caption{ Upper panel: the transverse response corresponding to the
  pure 2-body current is displayed (solid). Also shown are the
  contribution of the $\Delta$ current alone (thin-dashed) and of the
  seagull plus pion-in-flight only (thick-dashed).  Lower panel: the
  transverse response corresponding to the pure seagull (S, thin-dashed),
  pion-in-flight (P, thick-dashed) and S+P (solid) currents. }
\label{Fig4}
\end{center}
\end{figure}

The change of sign of the $\Delta$ contribution and the associated
bump in the response function are produced by the pion propagator.  In
the shell-model studies we use a dynamical pion propagator ({\it
  i.e.,} the exact one) which depends on the energy of the exchanged
pion.  The present computations are done in position space, and the
pion propagator is Fourier-Bessel transformed through a multipole
expansion (but is still energy-dependent).  For pion energies above
the pion mass a pole occurs in the Fourier integral, giving rise to a
change of sign in the $\Delta$ contribution (note that the pole is
treated as a principal value, hence no real pions are being produced).
Since in all of the MEC diagrams there is an exchanged pion, one
should also expect a bump in the seagull and pionic contributions.  In
fact those bumps are present, but small. This is better illustrated in
the example of Fig.~4, where we show for $q=1.5$ GeV/c the
1-body/2-body interference contributions to the T response.  There we
see that the total MEC contribution has the same oscillatory behavior
as the $\Delta$ current, contributing to the MEC bump in the high
$\omega$ tail in Fig.~2.  The largest contribution comes from the
$\Delta$ current, while the seagull plus pionic currents are very
small. On closer inspection of the lower panel we see that in fact
they both show a similar structure with the expected oscillation and
bump at high $\omega$.  The seagull and pionic contributions are
opposite in sign, and almost cancel out; accordingly, their net
contribution is small and the $\Delta$ dominates the MEC.

\begin{figure}[tph]
\begin{center}
\includegraphics[scale=0.6,  bb= 150 490 430 770]{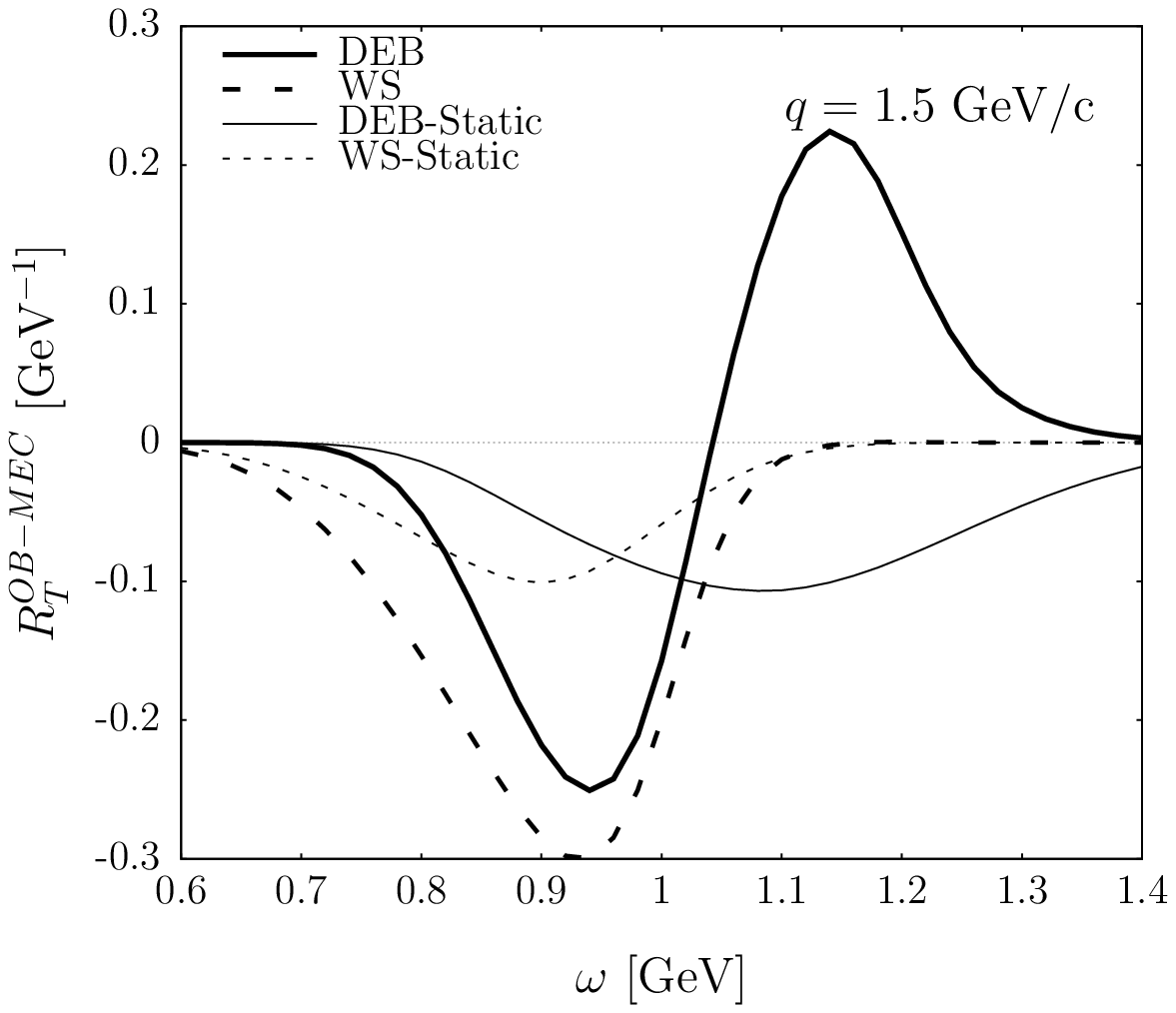}
\caption{ Interference 1-body/2-body transverse response obtained with
  the DEB (solid) and WS (dashed) potentials. The thick lines
  correspond to the exact dynamical pion propagator, the thin lines to
  the static approximation. }
\label{Fig5}
\end{center}
\end{figure}

Further insight into the enhancement of MEC from the pion dynamical
propagator is illustrated in Fig.~5, where we compare the full
calculation with the DEB results using a static pion propagator.  The
oscillation and bump disappear when the static pion propagator is used
and the MEC contribution is significantly reduced.  A similar effect
is also observed with a WS potential, even if here the MEC bump is
absent.  Concerning the effect of the DEB potential compared with the
WS results, we see that in both cases (static or dynamic pions) a
hardening of the response is observed, although in the dynamical case
an additional change of sign is produced.  This change of sign is
related to the oscillatory behavior of the pion propagator in
coordinate space for high pion energies, in contrast to its
exponential Yukawa-type behavior in the static case. 
Unfortunately, a simple estimate of the $\omega$ value
where the bump appears is not possible, since that value does not
depend on the kinematics in a trivial way.  In fact the pion
propagator appears inside an involved integration containing the
nuclear wave functions over an internal coordinate which is not
attached to the pion in the diagrams of Fig.~1. A more detailed
investigation of the physical origin and energy dependence of these
results is currently being pursued and will be reported elsewhere.

Summarizing, in the present work we have computed the MEC effects in
the transverse 1p-1h response for high momentum transfers in a
continuum SR shell model using the DEB potential for the final states.
The MEC are found to give an important contribution to the response,
which is negative at the peak and positive in the high-energy tail,
due to a change of sign of the MEC contribution for transferred
energies above 1 GeV.  These results confirm the MEC as an important
source of scaling violations in the T response at high $q$.  The MEC
bump is only predicted when the FSI produce a dynamical enhancement of
the high-energy tail of the responses (as in the case of the DEB
potential) and when this enhancement works in concert with the
dynamical pion propagator.

Although the effects shown in the present work are usually masked by
other contributions appearing at the kinematical region considered
($\Delta$ production, two-particle emission, {\it etc.}) and cannot be
separately observed in inclusive experiments, we have shown here that
they are not negligible.  This has important implications for neutrino
reaction studies \cite{Umi95}, 
where the contributions reported in the present work
must, of course, also be included.

\section*{Acknowledgements}
This work was partially supported by DGI (Spain):
FIS2008-01143, FPA2006-13807-C02-01, FIS2008-04189, FPA2007-62216,
by the Junta de Andaluc\'{\i}a, 
by the INFN-MEC collaboration agreement,
project ``Study of relativistic dynamcis in neutrino and
electron scattering'', the Spanish Consolider-Ingenio 2000
programmed CPAN (CSD2007-00042),
and part (TWD) by U.S. Department of Energy under cooperative
agreement DE-FC02-94ER40818.



\end{document}